\documentclass[aps,prd,a4paper,onecolumn,amsmath,showpacs,superscriptaddress,nofootinbib,preprintnumbers,notitlepage]{revtex4-1}
\usepackage{amssymb,amsmath}
\usepackage{graphicx}
\usepackage{color}
\usepackage{pgfplots}
\usepackage{booktabs}
\usepackage{multirow}
\usepackage{dcolumn}
\usepackage{amsmath}
\usepackage{mathtools}
\usepackage{amsfonts}
\usepackage{amssymb}
\usepackage{epstopdf}
\usepackage{bm}
\usepackage{braket}
\usepackage{enumitem}
\usepackage{soul}
\usepackage{ulem}
\usepackage{color}
\usepackage{transparent}
\usepackage{pifont}
\usepackage[font={small}]{caption}
\usepackage{float}
\usepackage{orcidlink}

\pgfplotsset{compat=1.18}

\usepackage{mathrsfs}
\begin{document}

\title{Reconstructing inflation  in Einstein-Gauss-Bonnet gravity in light of ACT data}

\author{Ram\'on Herrera\orcidlink{0000-0002-6841-1629}}
\email{ramon.herrera@pucv.cl}
\affiliation{Instituto de F\'{\i}sica, Pontificia Universidad Cat\'{o}lica de Valpara\'{\i}so, Avenida Brasil 2950, Casilla 4059, Valpara\'{\i}so, Chile.
}

\author{Carlos Ríos\orcidlink{0000-0003-3228-0003}}
\email{carlos.rios@ucn.cl}
\affiliation{Departamento de Ense\~nanza de las Ciencias B\'asicas, Universidad Cat\'olica del Norte, Larrondo 1281, Coquimbo, Chile.}

\begin{abstract}

During the inflationary epoch, we investigate the reconstruction of the background variables within the framework of Einstein-Gauss-Bonnet gravity, considering the scalar spectral index $n_s(N)$ and the tensor-to-scalar ratio $r(N)$, where $N$ denotes the number of $e-$folds. Under a general formalism, we determine the effective potential and the coupling function associated with the Gauss-Bonnet term as functions of the cosmological parameters $n_s(N)$ and $r(N)$, respectively. To implement the reconstruction  methodology for the background variables, we study an example in which the attractors for the index $n_s$
and the ratio $r$ are in agreement with  Atacama Cosmology Telescope (ACT) data. In this context, explicit expressions for the effective potential $V(\phi)$ and the coupling parameter $\xi(\phi)$ are reconstructed. Moreover, the reconstruction based on  observational parameters shows that $V(\phi)\not\propto 1/\xi(\phi)$, in contrast to the assumption 
 adopted in the literature for the study of the  evolution of the universe in Einstein-Gauss-Bonnet gravity.

\end{abstract}

\maketitle

\section{Introduction}

It is well known that, during its earliest scenario, the universe underwent an epoch of accelerated expansion, commonly referred to as cosmic inflation\cite{Starobinsky:1980te,Guth:1980zm,Linde:1981mu}. The inflationary scenario was  initially proposed as a mechanism to resolve different problems of the Big Bang model, including the flatness and horizon problems, among others. However, a crucial feature  of the inflationary epoch is its ability to provide a consistent explanation for the origin of the large-scale structure (LSS) of the universe, as well as for the primordial fluctuations that give rise to anisotropies in the cosmic microwave background (CMB)\cite{Mukhanov:1981xt,Hawking:1982cz,Guth:1982ec}. 

Among the different theoretical frameworks proposed to study
the inflationary era during the early evolution of the universe, particular attention has been devoted to models that include a Gauss-Bonnet correction. In the literature, there are numerous inflationary models within the framework of Einstein-Gauss-Bonnet gravity \cite{Carter:2005fu,Odintsov:2020sqy,
Kanti:2015pda,Guo:2010jr,Lidsey:2003sj,Pozdeeva:2024kzb,Oikonomou:2025ccs,Oikonomou:2024jqv}. In this context, string theory provides a compelling framework for achieving a unified description of gravity and the remaining fundamental interactions, and parallel theory offers an adequate approach to quantum gravity. In the low-energy scale, superstring theories give rise to effective actions that incorporate higher-curvature corrections. These corrections or additional terms  can become important in the high-energy regime of the early universe, and among them, the Gauss-Bonnet (GB) invariant represents the leading correction in the string effective actions, see e.g.
\cite{Gross:1986mw,Clifton:2011jh,Gasperini:1996fu}. The GB term corresponds to the leading higher-order curvature correction to the gravitational action derived from string theory frameworks. In the context of string theory, the GB term is naturally coupled to a scalar field $\phi$ from a function $\xi(\phi)$, see Refs.\cite{Nojiri:2005vv,Nojiri:2007te,Satoh:2007gn}.  Generally, this coupling function $\xi(\phi)$ is chosen  to be inversely proportional to the effective potential $V(\phi)$, during the early and late universe, since this assumption  enables the derivation of analytical solutions to the equations of motion; see e.g., Refs. \cite{Nojiri:2005vv,Guo:2010jr,Pozdeeva:2024kzb,Oikonomou:2025ccs,Oikonomou:2024jqv}.   

On the other hand, the reconstruction of the effective potential driving the dynamics of cold inflation considering the methodology derived from the observational quantities, such as; the scalar power spectrum, the scalar spectral index, and the tensor-to-scalar ratio has been extensively studied in the literature, see e.g., \cite{Hodges:1990bf,Easther:1995pc,Martin:2000ei,Herrera:2015udk,Belinchon:2019nhg,Li:2003ct}. In this context, we note that 
the methodology for reconstructing the background dynamical variables (e.g. the effective potential) in the case of a single scalar field, assuming a primordial scalar power spectrum, was first developed in Ref.\cite{Hodges:1990bf}. Here, the slow roll approximation was used without assuming any specific functional expression for the scalar spectral index. An interesting methodology for reconstructing the background dynamical variables under the slow roll approximation is formulated from  parameterization in terms of the number of $e-$folds $N$ of certain observational cosmological  parameters, such as the scalar spectral index, the tensor-to-scalar ratio or other. In particular, observational constraints from Planck data \cite{Planck:2018vyg} are in good agreement with the parametrization of the index $n_s\simeq 1-2/N$ and the tensor-to-scalar ratio $r\propto 1/N^{2}$, considering that the number of $e-$folds $N$ during the inflationary scenario is contained in the range $50<N<70$. In particular, the value of the scalar spectral index reported for the Planck 2018 data is  
$n_s=0.9649\pm0.0042$ \cite{Planck:2018vyg}.
Taking into account, large values of the number of $e-$folds $N$ ($N\gg 1$), the spectral index $n_s\simeq 1-2/N$, along with various expressions for the ratio $r(N)$, can be derived from a range of inflationary models. These models include the T-model \cite{Kallosh:2013hoa}, the E-model \cite{Kallosh:2013maa}, the Starobinsky $R^2-$model \cite{Starobinsky:1980te}, among others within the framework of General Relativity; see Ref.\cite{Chiba:2015zpa}. In other inflationary  models and  theories, different reconstructions of the background variables have been realized. For example, in warm inflation the reconstructions of the potential and the dissipation coefficient as functions of the scalar field using $n_s(N)$ and $r(N)$ were developed in the strong and weak regimes, respectively \cite{Herrera:2018cgi}. In the context of the Galileon model, the reconstruction using the  attractors $n_s(N)$ and $r(N)$ was studied in Ref.\cite{Herrera:2018mvo}.  Recently, the reconstruction of the background variables from observational parameters in the framework of a generalized  Rastall theory of gravity  was analyzed in Ref.\cite{Herrera:2025duu}, see also Refs.\cite{Herrera:2019xhs,Herrera:2020mjh,Herrera:2022kes,Gonzalez-Espinoza:2021qnv,Herrera:2023mas,Herrera:2024ojo,Herrera:2024aqf}.

However, observations from the ACT \cite{AtacamaCosmologyTelescope:2025blo,AtacamaCosmologyTelescope:2025nti} in conjunction with other data, indicate a higher value for the scalar spectral index compared to the results obtained by Planck. The ACT data report a value of the scalar spectral index $n_s=0.9743\pm0.0034$ \cite{AtacamaCosmologyTelescope:2025blo,AtacamaCosmologyTelescope:2025nti}, where  the analysis includes  baryon acoustic oscillation (BAO) measurements \cite{DESI:2024uvr}. Thus, recent observational data suggest that the scalar spectral index $n_s=1-2/N$ is disfavored
 at approximately $2\sigma$ level, when this expression  is evaluated at  $N=60$.  In this form, this result gives rise to a tension within a broad class of models that predict them.  Consequently, several inflationary models have been recently proposed to explain the observed increase in the value of the scalar spectral index $n_s$. In particular, as shown in Ref.\cite{Kallosh:2025rni}, the simplest
generalization of the chaotic model in which the effective potential $V(\phi)\propto\phi^2$, within the framework of a coupling to gravity $(1+\phi)R$, provides a good fit to the ACT data. In this model, the dependence of the  spectral index as a function of the number of $e-$folds $N$ is given by   $n_s=1-3/2N$. In the context of GB gravity \cite{Zhu:2025twm}, the increase in the value of $n_s$ can be alleviated for different effective potentials $V(\phi)$,  particularly when the coupling function $\xi(\phi)$ is defined as $\xi(\phi)\propto 1/V(\phi)$. In this sense, the GB coupling can give a wide class of models and improved consistency with the observational data, see also Ref.\cite{Yogesh:2025wak}. For a review of alternative inflationary models in light of ACT data, see e.g., Refs.\cite{Liu:2025qca,Gialamas:2025kef,Gao:2025onc,Addazi:2025qra}

This article aims to reconstruct an inflationary scenario, specifically the background dynamical variables within the framework of Einstein-Gauss-Bonnet gravity by considering parameterizations of key cosmological observables, namely the scalar spectral index and the tensor-to-scalar ratio, in terms of the number of $e-$folds $N$. In this sense, we investigate how the background dynamics and the cosmological perturbations inherent to GB gravity affect the reconstruction of both;  the effective potential $V(\phi)$ and the coupling function associated with the GB term $\xi(\phi)$, by assuming attractor solutions consistent  with ACT data and constrained by them. Within a general formalism, we reconstruct the effective potential and the GB coupling function from the attractors $n_s(N)$ and $r(N)$, in accordance with ACT results. The reconstruction of the background dynamical variables is performed within the framework of Einstein-Gauss-Bonnet gravity under the slow-roll approximation.

In order to obtain analytical expressions for the background variables, we consider a specific example for the observational parameters $n_s(N)$  and $r(N)$. In particular, we will consider the attractor forms $n_s=1-\gamma/N$ and $r=1/(N(1+\beta N)^p)$ where $\gamma$, $\beta$ and $p$ are constants. Here, the specific value  $\gamma=3/2$ in $N=60$ is well corroborated by the ACT data. Thus, we will reconstruct the effective potential and coupling coefficient associated with the GB term as functions of the scalar field. This will be done in conjunction with the different constraints on the relevant parameters, as determined by observational data, particularly the results from the ACT.

The outline of the article is as follows: the next section shows a brief review of the background and perturbations levels within the framework of Einstein-Gauss-Bonnet gravity. In Sect.\ref{Rec}, we analyze the reconstruction of the background variables within the framework of  GB gravity under a general formalism. Here we find explicit expressions in integral form  for the effective potential and the coupling function associated with the Gauss-Bonnet term as functions of the number of $e-$folds $N$. In Sect.\ref{Ex}  we consider a concrete example to apply the methodology utilized. In this respect, we assume the attractors for the scalar spectral index and the tensor-to-scalar ratio, to reconstruct the effective potential $V(\phi)$ and the coupling parameter $\xi(\phi)$ associated
with the Gauss-Bonnet function. In Sect.\ref{Conc} we present our conclusions. In this article, we chose units so that $c=\hbar=8\pi G=1$.

\section{Einstein-Gauss-Bonnet gravity}\label{EGB}

In this section, we will present a brief analysis of the implications of considering the reconstruction of inflation in the context of  Einstein-Gauss-Bonnet gravity. To describe this theory, we consider that the action $S$ in this framework  is defined as 
\begin{equation}
    S=\frac{1}{2}\int \sqrt{-g}\,d^{4}x\left[R-g^{\mu\nu
    }\partial_\mu\phi\partial_\nu\phi-2V(\phi) -\xi(\phi)\mathcal{G}\right],  \end{equation}
where 
the quantity $g$ is the determinant of the metric $g_{\mu\nu}$, $R$ corresponds to the  Ricci scalar, the two terms $\frac{1}{2}g^{\mu\nu} \partial_{\mu}{\phi}\partial_{\mu}{\phi}$ and 
$V(\phi)$ denote the kinetic energy density and   the effective potential associated with the scalar field $\phi$, respectively. In addition, the function
$\mathcal{G}$ corresponds to the called Gauss-Bonnet invariant which  is a pure topological term in four dimensions and is defined as 
\begin{equation}
\mathcal{G}=R^2-4R_{\mu\nu}R^{\mu\nu}+R_{\mu\nu\rho\sigma}R^{\mu\nu\rho\sigma},
\end{equation}
and  $\xi(\phi)$ denotes  the coupling function between the scalar field and the Gauss-Bonnet term \cite{Nojiri:2005vv,Nojiri:2007te,Satoh:2007gn}.

In order to describe the early universe, we consider that the metric corresponds to a spatially flat Friedmann-Roberson-Walker (FRW), and the scalar field is described by a homogeneous scalar field. In this way, within the framework of Gauss-Bonnet gravity, the equations of motion are given by \cite{Guo:2010jr}

\begin{equation}
6H^2=\dot{\phi}^2+2V+24\dot{\xi}H^3,
\end{equation}
\begin{equation}
2\dot{H}=-\dot{\phi}^2+4\ddot{\xi}H^2+4\dot{\xi}H(2\dot{H}-H^2),\,\,\mbox{and}
\end{equation}
\begin{equation}
\ddot{\phi}+3H\dot{\phi}+V_\phi+12\xi_\phi H^2(\dot{H}+H^2)=0,
\end{equation}
respectively. Here, $H=\dot{a}/a$ denotes the Hubble parameter and the quantity $a(t)$ corresponds to the scale factor. The dots denote derivatives with respect to the cosmic time and the notation $V_\phi$ corresponds to $V_\phi=d V/d\phi$, $\xi_\phi=d \xi/d\phi$,  $V_{\phi\phi}=d^2 V/d\phi^2$, etc. 

Following Ref.\cite{Guo:2010jr}, we can consider that under the slow-roll approximation, both the scalar field $\phi$ and the coupling function $\xi(\phi)$ vary slowly. In this form, we can introduce the slow-roll conditions in the framework of Gauss-Bonnet gravity as  
\begin{equation}
|\ddot{\phi}|\ll 3H|\dot{\phi}|,\,\,\,\,\,\,\,\,\dot{\phi}^2\ll V,\,\,\,\,\,\,\,\,\,4|\dot{\xi}|H\ll1,\,\,\,\,\,\,\,\mbox{and}\,\,\,\,\,\,\,\,\,|\ddot{\xi}|\ll |\dot{\xi}|H.
\end{equation}
From these conditions, we introduce the Hubble flow parameters  $\epsilon_i$ defined as \cite{Guo:2010jr}
\begin{equation}
\epsilon_{1}=-\frac{\dot{H}}{H^{2}}, \quad \epsilon_{i+1}=\frac{d\ln \epsilon_{i}}{d \ln a},\quad i\geq1, \quad \mbox{e.g., if $i$=1, then },\,\,\,\,\quad \epsilon_2=-\frac{1}{\epsilon_1}\frac{d\epsilon_1}{dN},\label{E1}
\end{equation}
where we have used that the relation between the number of $e-$folds $N$ and the scale factor is given by the relation $dN=-H\,dt=-d\ln\,a$. In this context, these parameters correspond to the conventional slow roll parameters. Thus,  the inflationary era of the universe occurs when $\ddot{a}>0$, which implies that the parameter $\epsilon_1<1$. In addition, the end of the inflationary epoch occurs  when the parameter $\epsilon_1=1$ or equivalently when $\ddot{a}=0$.

In relation to the additional degrees of freedom related to the Gauss Bonnet coupling function $\xi(\phi)$, we can introduce a new hierarchy of flow parameters, defined analogously by \cite{Guo:2010jr,Schwarz:2001vv}
\begin{equation}
\delta_{1}=4\dot{\xi}H,\quad \delta_{i+1}=\frac{d \ln \delta_{i}}{d\ln a}, \quad i\geq1, \quad \Rightarrow \quad \delta_2=-\frac{1}{\delta_1}\frac{d\delta_1}{dN}.\label{D1} 
\end{equation}
Therefore, in the framework of Gauss-Bonnet gravity, the slow-roll approximation is characterized by the conditions $|\epsilon_i|\ll1$ and $|\delta_i|\ll 1$.

Thus, assuming these slow-roll conditions, we can write that the equations of motion are reduced to \cite{Guo:2010jr}
\begin{equation}
H^2\simeq\frac{1}{3}\,V,\label{H1}
\end{equation}
\begin{equation}
\dot{H}\simeq-\frac{1}{2}\dot{\phi}^2-2\dot{\xi}H^3,
\end{equation}
and 
\begin{equation}
3H\dot{\phi}\simeq-(V_\phi+12\xi_\phi H^4).\label{3H}
\end{equation}
We note that, in the context of the slow-roll approximation,  the above  equations reduce to the General Relativity case in the limit when the GB parameter $\xi\rightarrow $0.

In this framework, the duration of inflation  characterized by the number of $e-$folds $N$  becomes
\begin{equation}
\label{deltaN}
    \Delta N=N-N_\text{end}=-\int^{t}_{t_\text{end} }H\,dt=-\int^{\phi}_{\phi_\text{end}}\left(\frac{H}{\dot{\phi}}\right)d\phi\simeq\int^{\phi}_{\phi_\text{end}}\,\frac{d\phi}{Q},
\end{equation}
where the quantities $N_\text{end}$ and $t_\text{end}$ (or $\phi_\text{end}$) correspond to the values of the number of $e-$folds and the time (or scalar field) at which the inflationary epoch ends. Furthermore,  the function $Q$ associated to the potential and the coupling function $\xi$, is defined as 
\begin{equation}
Q=\frac{V_\phi}{V}+\frac{4}{3}\xi_\phi V,\label{Q}
\end{equation}
where, we have used Eqs.(\ref{H1}) and (\ref{3H}), respectively.

On the other hand, in the framework of Einstein-Gauss-Bonnet gravity, the power spectrum of the curvature perturbations and the scalar  spectral index were determined in Refs.\cite{Guo:2010jr,Yi:2018gse}. In this respect, the Mukhanov-Sasaki equation  associated with the scalar perturbation in this framework can be written as 
\begin{equation}
v_k''+\left(c_s^2k^2-\frac{z_s''}{z_s}\right)v_k=0,\label{Mk}
\end{equation}
where a prime corresponds to the derivative with respect  to the conformal time $\tau=\int a^{-1}dt$ and $k$ denotes the wave number. The effective sound speed $c_s$  and the function  $z_s$ are defined as \cite{Guo:2010jr,Yi:2018gse}
\begin{equation}
c_s^2=1-\Delta^2\left[\frac{2\epsilon_1+\delta_1(1-5\epsilon_1-\delta_2)/2}{F}\right],\,\,\,\,\,\mbox{and}\,\,\,\,\,\,\,\,z_s^2=a^2\left[\frac{F}{(1-\Delta/2)^2}\right],
\end{equation}
respectively. Here, the auxiliary functions $\Delta$ and $F$ associated with the slow-roll parameters  are defined as
\begin{equation}
\Delta=\frac{\delta_1}{1-\delta_1}\,\,\,\,\,\,\mbox{and}\,\,\,\,\,\,\,\,F=2\epsilon_1-\delta_1(1+\epsilon_1-\delta_2)+3\delta_1\Delta/2.\label{FF}
\end{equation}
Thus, following Refs.\cite{Guo:2010jr,Yi:2018gse}, we can write that the effective mass term $z_s''/z_s$ associated to Eq.(\ref{Mk}) as a function of the conformal time $\tau$ becomes
\begin{equation}
\frac{z_s''}{z_s}=\frac{1}{\tau^2}\left(\nu-\frac{1}{4}\right),
\end{equation}
where the parameter $\nu$ related to the slow roll parameters is defined as
\begin{equation}
    \nu=\frac{3}{2}+\epsilon_1+\frac{2\epsilon_1\epsilon_2-\delta_1\delta_2}{2(2\epsilon_1-\delta_1)}\label{nu}.
\end{equation}
In this way, the power spectrum for the scalar perturbation $A_s$ on superhorizon scales can be written as  \cite{Guo:2010jr,Yi:2018gse}
$$
    A_s=\frac{k^3}{2\pi^2}\left|\frac{v_k}{z}\right|^2=2^{2\nu-3}\left[\frac{\Gamma(\nu)}{\Gamma(3/2)}\right]^2 (1-\epsilon_1)^{2\nu-1} \left. \frac{(1-\frac{1}{2}\Delta)^2}{Fc_s^3}\left(\frac{H}{2\pi}\right)^2 \left(\frac{c_s k}{aH}\right)^{3-2\nu}\right|_{c_s k=aH}   
$$   
 \begin{equation}    
    \simeq\frac{2^{2\nu-3}}{Fc_s^3}\left[\frac{\Gamma(\nu)}{\Gamma(3/2)}\right]^2  \left. \left(\frac{H}{2\pi}\right)^2 \right|_{c_s k=aH},\label{As1}
\end{equation}
where the function $\Gamma$ corresponds to the Gamma function, and the power spectrum is evaluated at the horizon crossing, i.e.,  when $c_sk=aH$.

From the power spectrum given by Eq.(\ref{As1}), the scalar spectral index $n_s$ as a function  of the slow roll parameters results \cite{Guo:2010jr,Yi:2018gse}
\begin{equation}  
n_s-1=\frac{d\ln A_s}{d\ln k}=3-2\nu=-2\epsilon_1-\left[\frac{2\epsilon_1\epsilon_2-\delta_1\delta_2}{2\epsilon_1-\delta_1}\right],\label{ns1}
\end{equation}  
where we have considered the value of $\nu$ defined by Eq.(\ref{nu}).

Similarly, the  Mukhanov-Sasaki equation for  tensor perturbations in the framework of Gauss-Bonnet gravity can be written as  \cite{Guo:2010jr,Yi:2018gse}
\begin{equation}  
u_k''\,^\lambda+\left(c_t^2k^2-\frac{z_t''}{z_t}\right)u_k^\lambda=0,
\end{equation}  
where superscript  $\lambda$ denotes   ``+'' o ``x'' polarizations and the quantities $c_t$ and $z_t$ are defined as
$$
c_t^2=1+\Delta(1-\epsilon_1-\delta_2),\,\,\,\,\,\mbox{and}\,\,\,\,\,\,\,z_t^2=a^2(1-\delta_1),
$$
respectively. 

In this way, the power spectrum $A_t$ related to the tensor perturbations in Gauss-Bonnet gravity becomes  \cite{Guo:2010jr,Yi:2018gse}
\begin{equation}  
A_t=\left[\frac{2^{2u}\,(1-\epsilon_1)^{2u-1}}{(1-\delta_1)c_t^3}\right]\,\left[\frac{\Gamma(u)}{\Gamma(3/2)}\right]\,\left(\frac{H}{2\pi}\right)^2\,\left.\left(\frac{c_tk}{aH}\right)^{3-2u}\right|_{c_s k=aH},\label{At}
\end{equation}  
where the parameter $u$ is defined as $u=3/2+\epsilon_1$, and as before this  spectrum is evaluated at the horizon crossing.

In addition, we can consider a fundamental observational parameter called the tensor-to-scalar ratio $r$, which is defined as the ratio between tensor and scalar perturbations. Within the   Gauss-Bonnet framework, the ratio $r$ is expressed in terms of the flow parameters $\epsilon_1$ and $\delta_1$ as  \cite{Guo:2010jr,Yi:2018gse}
\begin{equation}
    r=16\epsilon_1-8\delta_1.\label{r1}
\end{equation}

Additionally, from Eq.(\ref{At}) it is possible to determine that the tensor spectral index $n_t$ is given by
\begin{equation}
n_t=\frac{d\ln A_t}{d\ln k}=3-2u=-2\epsilon_1.
\end{equation}

We note that in  Gauss-Bonnet gravity, the consistency relation is modified and  does not satisfy the standard expression within the framework of General Relativity, where the ratio $r$ is defined as  $r=-8n_t$.

\section{Reconstructing inflation in Gauss-Bonnet gravity: General reconstruction from $n_s(N)$ and $r(N)$}\label{Rec}

In this section, we make use of the methodology to reconstruct
the background variables from the observational parameters. In this sense,  we will rebuild the effective scalar potential $V(\phi)$ and the coupling function $\xi(\phi)$ in terms of the scalar field, considering the parameterization of the scalar spectral index and the tensor to scalar ratio as  functions of the number of $e-$folds $N$. Hence, the  reconstruction involves two background variables associated with the scalar field; the potential $V(\phi)$ and the coupling function $\xi(\phi)$, then the reconstruction procedure consists of employing two observational parameters expressed in terms of the number of $e-$folds $N$, which are; $n_s(N)$ and $r(N)$. In this context, we need to rewrite the scalar spectral index and the tensor-to-scalar ratio given by Eqs.(\ref{ns1}) and (\ref{r1}) as functions of the potential $V$ and the coupling parameter  $\xi$ in terms of the number of $e-$folds $N$ and its derivatives. In this form, from these relations and specifying the observational parameters $n_s(N)$ and $r(N)$, we should determine the effective potential $V(N)$ and the coupling parameter $\xi(N)$. Subsequently, using Eq.(\ref{deltaN})  we should obtain the number of $e-$folds $N$ as a function of the scalar field i.e., $N=N(\phi)$ to finally reconstruct the effective potential $V(\phi)$ and the coupling function $\xi(\phi)$, respectively.

We begin by rewriting the scalar spectral index and the tensor to scalar ratio defined by Eqs.(\ref{ns1}) and (\ref{r1}) as functions of the number of $e-$folds $N$. For the scalar spectral index, we consider that the last term in Eq.(\ref{ns1}) can be written as 
\begin{equation}
 \quad 2\epsilon_1\epsilon_2-\delta_1\delta_2=\frac{d\delta_1}{dN}-2\frac{d\epsilon_1}{dN}=-\frac{1}{8}\frac{dr}{dN}\,\,\,\,\,\,\Rightarrow\,\,\,\,\,\,\, \frac{2\epsilon_1\epsilon_2-\delta_1\delta_2}{2\epsilon_1-\delta_1}=-\frac{1}{r}\frac{dr}{dN}=-\frac{d\ln r}{dN},
\end{equation}
where we have used $2\epsilon_1-\delta_1=r/8$ together with Eqs.(\ref{E1}) and (\ref{D1}), respectively. Now, under the slow roll approximation and following Refs.\cite{Guo:2010jr,Yi:2018gse}, the parameter $\epsilon_1\simeq QV_\phi/(2V)$ can be rewritten as 
\begin{equation}
\epsilon_1\simeq\frac{Q}{2}\frac{V_{\phi}}{V}=\frac{1}{2}\frac{V_N}{V},\label{epsi1}
\end{equation}
where we have considered  $dN=d\phi/Q$. In the following, we will consider that  the subscription $V_N=dV/dN$, $V_{NN}$ corresponds to $V_{NN}=d^2V/dN^2$, $\xi_N=d\xi/dN$, etc. This allows us to rewrite Eq.(\ref{ns1}) as follows
\begin{equation}
    n_s-1=\frac{d\ln r}{dN}-\frac{d\ln V}{dN}=\left[\ln\left(\frac{r}{V}\right)\right]_N.\label{dif1}
\end{equation}
In this form, solving Eq.(\ref{dif1}) in terms of the number of $e$-folds $N$, we find that the effective potential $V(N)$ results
\begin{equation}
    V(N)=r(N)\exp\left[-\int\left(n_s-1\right)dN\right].\label{V(N)}
\end{equation}
Here we note that to obtain the effective potential as a function of the number of $e-$folds $N$, we need to specify the observational parameters $n_s(N)$ and $r(N)$, respectively.  From this solution, we determine that the ratio $V_N/V$ in terms of the observational parameter and its derivatives can be written as
\begin{equation}
\frac{V_N}{V}=\left[(1-n_s)+\frac{r_N}{r}\right].
\end{equation}
Thus, using the expression for the tensor-to-scalar ratio given by Eq.(\ref{r1}) yields 
\begin{equation}
r=8\left[\frac{V_N}{V}+\frac{4}{3}\xi_N V\right]\,\,\,\,\,\Rightarrow\,\,\,\,\,\,\,\xi_N=\left[\frac{r}{8}-\frac{V_N}{V}\right]\frac{3}{4V},\label{E1}
\end{equation}
where we have utilized that the slow-roll parameter under the slow roll approximation is given by $\delta_1=-4\xi_\phi QV/3$ \cite{Guo:2010jr,Yi:2018gse}.
In this way, by integrating Eq.(\ref{E1}), we find that the coupling parameter $\xi(N)$ as a function of the number of $e-$folds $N$ can be determined from the integral expression
\begin{equation}
\xi(N)=\frac{3}{4}\int\left\{\frac{1}{r}\Bigg[\frac{r}{8}+(n_s-1)-\frac{r_N}{r}\Bigg]\exp\left[\int\left(n_s-1\right)dN\right]\right\}dN.
\label{xi}\end{equation}
As before, to determine the coupling parameter $\xi(N)$, we need to specify the observational parameters in terms of the number of $e-$folds.  

Finally, to obtain the relation between the number of $e-$folds $N$ and the scalar field $\phi$, we consider Eqs.(\ref{deltaN}) and (\ref{Q}), such that 
\begin{equation}
Q\,dN=\sqrt{\left[\frac{V_N}{V}+\frac{4}{3}\xi_N\,V\right]}\,\,dN=d\phi.\label{Nf}
\end{equation}
Here, we note that in the  GR limit, where the parameter $\xi\rightarrow 0$, Eq.(\ref{Nf}) reduces to the standard expression for $N=N(\phi)$, given by the expression $\sqrt{V_N/V}\,dN=d\phi$, see Ref.\cite{Chiba:2015zpa}.

Thus, Eqs.(\ref{V(N)}) and  (\ref{xi}) together with Eq.(\ref{Nf}) are the fundamental equations to reconstruct the effective potential $V(\phi)$ and the coupling parameter $\xi(\phi)$, as functions of the scalar field $\phi$, from the attractors $n_s(N)$ and $r(N)$.

In the following, we implement our reconstruction methodology of the background variables to a specific example of the observational parameters $n_s(N)$ and $r(N)$.

\subsection{An Example}\label{Ex}

In this section, we apply the above methodology within  the framework  of Einstein-Gauss-Bonnet gravity, using as attractors  the scalar spectral index and the tensor-to-scalar ratio  as  functions of the number of $e-$folds $N$,
 to analytically reconstruct   the effective potential
$V(\phi)$ and the coupling parameter $\xi(\phi)$ in terms of the scalar field during the inflationary epoch. 

 Regarding  the scalar spectral index as a function of the number of  $e-$folds $N$ and following Refs.\cite{Starobinsky:1980te,Chiba:2015zpa,Kallosh:2013hoa}, 
we consider that this index is given by 

\begin{equation}
  n_s(N)=  n_s=1-\frac{\gamma}{N},\label{nsAtr}
\end{equation}
where $\gamma$ is a positive constant and nearly equal to two. In particular,  using the observational data reported by  ACT in which the scalar spectral index $n_s=0.974$ and considering that the number of $e-$folds  $N=60$, then the constant $\gamma=1.56\simeq3/2$. In this way,  the relation given by Eq.(\ref{nsAtr}) can be interpreted as a  generalization of the attractor for the scalar spectral index defined as $n_s=1-2/N$.

Analogously, for the parameterization of the tensor-to-scalar ratio $r$ as a function of the number of $e-$folds $N$, we assume a generalization of the  $\alpha$ attractor scenario \cite{Kallosh:2013yoa,Jinno:2017jxc},  as well as  the parameterization $r=(N[1+\beta N])^{-1}$ \cite{Herrera:2018cgi}. In this sense, we can consider that the tensor-to-scalar ratio can be written as  

\begin{equation}
    r(N)=r=\frac{1}{N\left(1+\beta N\right)^p},\label{rAtr}
\end{equation}
where, for simplicity, we assume that the exponent $p$ is a constant and  a positive integer ($p=1,2,3,..$) and that $\beta$ is also  a constant. In particular, for the special case in which $p=1$, Eq.(\ref{rAtr}) reduces to the ratio  $r(N)$ studied in Ref.\cite{Herrera:2018cgi} and for  $\beta N\gg 1$, this tensor-to-scalar  ratio corresponds to the $\alpha-$attractor scenario,  where $r=1/(\beta N^2)$ with $\beta=1/(12\alpha)$, see Refs.\cite{Kallosh:2013yoa,Jinno:2017jxc}.

Regarding Eqs.(\ref{nsAtr}) and (\ref{rAtr}), if we assume that  when the number of $e-$folds at the horizon exit is in the range $N\simeq50-70$, then the scalar spectral index evaluated in the range $n_s(N\simeq50-70)$ is well corroborated by observation data, see, e.g., Refs.\cite{AtacamaCosmologyTelescope:2025blo,AtacamaCosmologyTelescope:2025nti,Planck:2018vyg}. For instance, for the case $\gamma=3/2$ in  $N=60$, the scalar spectral index defined by  Eq.(\ref{nsAtr}) is well supported by the ACT data, in which $n_s\simeq0.97$.
In addition, we note that  for the tensor-to-scalar ratio,  we require that for  odd values of $p$, the parameter $\beta$ satisfies the condition $\beta>-1/N$, since the ratio $r$ is a positive-definite quantity. In addition, for fixed values of the exponent $p$ and $N$, the parameter $\beta$ is given by the expression $\beta=[(rN)^{-1/p}-1]/N$. In this way, if we consider the upper limit for the tensor to scalar ratio $r=0.039$ (at 1-$\sigma$ confidence level), see  Ref.\cite{Planck:2018vyg} and the number $N=60$, then for $p=1$ we have $\beta\simeq-0.0095$, if $p=2$ then $\beta\simeq-0.0058$, if $p=3$ results $\beta\simeq -0.0041$, etc. In addition, considering  the special case  $1/N>r$, we have that the parameter $\beta$ is a positive define quantity.

In relation to the effective potential $V(N)$ as a function of the number of $e-$folds $N$ defined by Eq.(\ref{V(N)}), we find that using the attractor given by Eq.(\ref{nsAtr}), the term $\exp\int (1-n_s)dN=N^\gamma\alpha$, where the parameter $\alpha$ is associated to an integration constant.
In this form, replacing  Eq.(\ref{rAtr}) into Eq.(\ref{V(N)}), we find that the effective potential $V(N)$ becomes
\begin{equation}
V(N)=\frac{\alpha\, N^{\gamma-1}}{\left(1+\beta N\right)^p}.\label{VNN}
\end{equation}
Since the effective potential $V(N)$ is a positive-definite  quantity, we require  the integration constant $\alpha>0$, for every even integer $p$ when $\beta \lessgtr 0
$. If the power $p$ is odd and $\beta<0$, then the parameter $\alpha$ is positive provided that $1>\beta N$. 

In order to obtain the coupling parameter $\xi$ in terms of the number of $e-$folds $N$, we make use of  Eqs.(\ref{xi}), (\ref{nsAtr}) and (\ref{rAtr}), obtaining

\begin{equation}
    \xi(N)=\frac{3}{4\alpha N^{\gamma-1}}\left[(1+\beta N)^p-\frac{1}{8(\gamma-1)}\right]+C,\label{xi1}
\end{equation}
where the quantity $C$ corresponds to  a new integration constant. In particular, assuming that the parameter $\gamma=2$, Eq.(\ref{xi1}) is reduced to
\begin{equation}   \xi(N)=\frac{3}{4\alpha N}\left[\left(1+\beta N\right)^p-\frac{1}{8}\right]+C.\label{xi2}
\end{equation}

In addition, the function $Q$ associated with the potential and the coupling function given by Eq.(\ref{Q}) can be written in terms of the number of $e-$folds $N$ as

\begin{equation}
    Q(N)=\sqrt{\frac{V_N}{V}+\frac{4}{3}\xi_N V}\,
    =\big[8N\left(1+\beta N\right)^p\big]^{-1/2},\label{Q2}
\end{equation}

where we have considered that the term $\xi_N$, from Eq.(\ref{E1}), is given by 
\begin{equation}
     \xi_N= -\frac{3}{4\alpha N^\gamma}\left\{\left[\gamma-1 +\beta  N \left(\gamma-1-p\right)\right]\left(1+\beta  N\right)^{p-1}-1/8\right\}.\label{xiN}
\end{equation}
In addition, we note that the parameter $Q(N)$ defined by Eq.(\ref{Q2}) does not depend on the parameter $\gamma$ associated to the attractor $n_s$ given by Eq.(\ref{nsAtr}).


In this way, from Eqs.(\ref{Nf}) and (\ref{Q2}), we find that the relation between the number of $e-$folds $N$ and the scalar field becomes
\begin{equation}
dN=\sqrt{8N\left(1+\beta N\right)^p}\,d\phi.\label{dnfp}
\end{equation}
Here, we note that the relation between the number of $e-$folds $N$ and the scalar field does not depend on the parameter $\gamma$ associated to the scalar spectral index $n_s$. 
In this way, the solution of the differential equation given by Eq.(\ref{dnfp}) depends on the sign of the constant $\beta$.  Thus, the different solutions will be described for the cases where the parameter $\beta$ takes the values  $\beta>0$ and $\beta<0$, respectively. The special case in which $\beta=0$ is not of interest, since the tensor-to-scalar ratio does not depend on the parameters $p$ and $\beta$; as a result, the ratio $r$  takes the fixed value $r=1/N$ for a given value of the number $N$. 

In this form, the solution of Eq.(\ref{dnfp}),    for either positive or negative  values of the parameter $\beta$ ($\beta\gtrless 0$) can be written as
\begin{equation}
N(\phi)=\mathcal{F}^{-1}(\phi),\label{Nf3}
\end{equation}
where the quantity  $\mathcal{F}^{-1}$ denotes the inverse of the function $\mathcal{F}(\phi)$ associated with the scalar field given by 
\begin{equation}
   \mathcal{F}(\phi)=
   2 \left(2\sqrt{2}\phi +C_p\right)^{1/2}  \, _2F_1\left[\,\frac{1}{2}\,,\,\frac{p}{2}\,;\,\frac{3}{2}\,;\,\pm\,|\beta|  \left(2\sqrt{2}\phi +C_p\right)\right],\label{sol1}
\end{equation}
where $C_p$ denotes different integration constants associated with parameter $p$ and  the function $_2F_1$ corresponds to the hypergeometric function. In addition, the negative sign in the expression  given by  Eq.(\ref{sol1}) is for the situation in which the parameter is $\beta>0$ while  the 
positive sign applies when $\beta<0$.

In particular,  for $p=1$, $p=2$ and  $p=3$ the different  solutions $N=N(\phi)$ given by Eq.(\ref{Nf3}) together with  Eq.(\ref{sol1}) when the parameter $\beta$ is defined positive, are given by 
\begin{align}
    p=1&\Rightarrow\quad N(\phi)=\frac{1}{\beta}\sinh^2\left(\sqrt{2\beta}\phi+C_1\right),\\
    p=2&\Rightarrow\quad N(\phi)=\frac{1}{\beta}\tan^2\left(\sqrt{2\beta}\phi+C_2\right),\,\,\,\,\,\,\,\,\,\,\,\mbox{and}\\
    p=3&\Rightarrow\quad N(\phi)=\frac{2\left(\phi+C_3\right)^2}{1-2\beta\left(\phi+C_3\right)^2}.
\end{align}

However, for the case in which the parameter $\beta$ is negative, the solutions  are given by 
\begin{align}
p=1 &\Rightarrow N(\phi)
= \frac{1}{|\beta|}\,
\sin^2\!\left(\sqrt{2|\beta|}\,\phi+C_1\right), \\
p=2 &\Rightarrow N(\phi)
= \frac{1}{|\beta|}\,
\tanh^2\!\left(\sqrt{2|\beta|}\,\phi+C_2\right), \,\,\,\,\,\,\,\mbox{and}\\
p=3 &\Rightarrow N(\phi)
= \frac{2\left(\phi+C_3\right)^2}
{1+2|\beta|\left(\phi+C_3\right)^2},
\end{align}
respectively. Here we have used that for $p=1$, the integration constant $C_{p=1}=C_1$, for $p=2$ corresponds to  $C_{p=2}=C_2$, etc.

In this form, combining Eqs.(\ref{VNN}) and (\ref{Nf3}), we obtain that the reconstruction of the effective potential as a function of the scalar field can be written as
\begin{equation}
V(\phi)=\frac{\alpha\,\left[\mathcal{F}^{-1}(\phi)\right]^{\gamma-1}}{[1+\beta\,\mathcal{F}^{-1}(\phi)]^p}.\label{Pot1}
\end{equation}

In particular, assuming $\beta>0$  together with the values of $p=1$, $p=2$ and $p=3$, we find that the corresponding  reconstructions  of the effective potentials in terms of the scalar field  are

\begin{align}
p=1 &\Rightarrow V(\phi)
=\alpha\,\text{sech}^2\left(\sqrt{2\beta}\phi+C_1\right)\left[
\frac{\sinh^2\!\left(\sqrt{2\beta}\phi+C_1\right)}{\beta}\right]^{\gamma-1}, \\
p=2 &\Rightarrow V(\phi)
=  \alpha\,\cos^4\left(\sqrt{2\beta}\phi+C_2\right)\left[
\frac{\tan^2\!\left(\sqrt{2\beta}\phi+C_2\right)}{\beta}\right]^{\gamma-1}, \,\,\,\,\,\,\mbox{and}\\
p=3 &\Rightarrow V(\phi)
= 2^{\gamma-1}\alpha\left[1-2\beta\left(\phi+C_3\right)^2\right]^3 \left[\frac{\left(\phi+C_3\right)^2}{1-2\beta\left(\phi+C_3\right)^2}\right]^{\gamma-1}.
\end{align}

Here we note that in the special case in which the parameter $\gamma$ associated with the scalar spectral index
is equal to two, the above  reconstructed effective potentials $V(\phi)$ are reduced to
\begin{align}
p=1 &\Rightarrow V(\phi)
= \left(\frac{\alpha}{\beta}\right)
\left[\frac{\sinh^2\!\left(\sqrt{2\beta}\phi+C_1\right)}
{1+\sinh^2\!\left(\sqrt{2\beta}\phi+C_1\right)}\right]
= \left(\frac{\alpha}{\beta}\right)
\tanh^2\!\left(\sqrt{2\beta}\phi+C_1\right), \\
p=2 &\Rightarrow V(\phi)
= \left(\frac{\alpha}{\beta}\right)
\frac{\tan^2\!\left(\sqrt{2\beta}\phi+C_2\right)}
{\left[1+\tan^2\!\left(\sqrt{2\beta}\phi+C_2\right)\right]^2}
= \left(\frac{\alpha}{\beta}\right)
\sin^2\!\left(\sqrt{2\beta}\phi+C_2\right)
\cos^2\!\left(\sqrt{2\beta}\phi+C_2\right), \,\mbox{and}\\
p=3 &\Rightarrow V(\phi)
= 2\alpha\left(\phi+C_3\right)^2
\left[1-2\beta\left(\phi+C_3\right)^2\right]^2 .
\end{align}

Now, if the parameter $\beta$ is considered a negative quantity, then the reconstruction of the effective potentials in terms of the scalar field for different values of the parameter $p$ result 
\begin{align}
p=1&\Rightarrow 
V(\phi)
=
\alpha\,|\beta|^{1-\gamma}\,
\frac{\sin^{2(\gamma-1)}\!\left(\sqrt{2|\beta|}\phi+C_1\right)}
{1+\sin^{2}\!\left(\sqrt{2|\beta|}\phi+C_1\right)},
\\[1.2em]
p=2&\Rightarrow
V(\phi)
=
\alpha\,|\beta|^{1-\gamma}\,
\frac{\tanh^{2(\gamma-1)}\!\left(\sqrt{2|\beta|}\phi+C_2\right)}
{\left[1+\tanh^{2}\!\left(\sqrt{2|\beta|}\phi+C_2\right)\right]^2},\,\,\,\,\,\,\,\,\,\,\,\,\,\mbox{and}
\\[1.2em]
p=3&\Rightarrow 
V(\phi)
=
\,2^{\gamma-1}\alpha\,\left[1+2|\beta|\,(\phi+C_3)^2\right]^3\left[\frac{(\phi+C_3)^2}{1+2|\beta|\,(\phi+C_3)^2}\right]^{\gamma-1}.
\end{align}

In particular, assuming $\beta<0$ and  the special case $\gamma=2$, the above reconstructed potentials are simplified to 

\begin{align}
p=1 &\Rightarrow V(\phi)
= \left(\frac{\alpha}{|\beta|}\right)
\left[\frac{\sin^2\!\left(\sqrt{2|\beta|}\phi+C_1\right)}
{1-\sin^2\!\left(\sqrt{2|\beta|}\phi+C_1\right)}\right]
= \left(\frac{\alpha}{|\beta|}\right)\tan^2\!\left(\sqrt{2|\beta|}\phi+C_1\right), \\
p=2 &\Rightarrow V(\phi)
= \left(\frac{\alpha}{|\beta|}\right)
\frac{\tanh^2\!\left(\sqrt{2|\beta|}\phi+C_2\right)}
{\left[1-\tanh^2\!\left(\sqrt{2|\beta|}\phi+C_2\right)\right]^2}
= \left(\frac{\alpha}{|\beta|}\right)
\sinh^2\!\left(\sqrt{2|\beta|}\phi+C_2\right)
\cosh^2\!\left(\sqrt{2|\beta|}\phi+C_2\right),
\end{align}
and 
\begin{align}
p=3 &\Rightarrow V(\phi)
= 2\alpha\left(\phi+C_3\right)^2
\left[1+2|\beta|\left(\phi+C_3\right)^2\right]^2 .
\end{align}

Analogously, using Eqs.(\ref{xi1}) and (\ref{Nf3}), we find that the reconstruction of the  coupling parameter associated with the Gauss-Bonnet term as a function  of the scalar field becomes

\begin{equation}
    \xi(\phi)=\frac{3}{4\alpha [\mathcal{F}^{-1}(\phi)]^{\gamma-1}}\left[(1+\beta \mathcal{F}^{-1}(\phi))^p-\frac{1}{8(\gamma-1)}\right]+C,\label{xiF}
\end{equation}
where $C$ corresponds to a new integration constant.

 We  also note that for the special situation in which  $\gamma=2$, the coupling parameter related to the Gauss-Bonnet term as a function of the scalar field given by Eq.(\ref{xiF}) is reduced to
\begin{equation}
\xi(\phi)=\frac{3}{4\alpha \mathcal{F}^{-1}(\phi)}\left[\left[1+\beta \mathcal{F}^{-1}(\phi)\right]^p-\frac{1}{8}\right]+C.\label{Xia}
\end{equation}

Thus, in particular  for  $\beta>0$  and $p=1,2,3$, the corresponding  reconstructions  of the coupling parameters associated with the Gauss-Bonnet term  as functions of the scalar field result

\begin{align}
p=1&\Rightarrow
\xi(\phi)
=
\frac{3}{32\alpha}\,
\beta^{\gamma-1}
\sinh^{2(1-\gamma)}\!\left(\sqrt{2\beta}\phi+C_1\right)\!
\Bigg[\!
\frac{1}{1-\gamma}\!+
\cosh^{2p}\!\left(\sqrt{2\beta}\phi+C_1\right)
\Bigg]\!
+\! C ,
\\[1.5em]
p=2&\Rightarrow 
\xi(\phi)
=
\frac{3}{32\alpha}\,
\beta^{\gamma-1}
\tan^{2(1-\gamma)}\!\left(\sqrt{2\beta}\phi+C_2\right)\!
\Bigg[\!
\frac{1}{1-\gamma}\!+
\sec^{2p}\!\left(\sqrt{2\beta}\phi+C_2\right)
\Bigg]\!
+\! C, \,\,\,\,\,\,\,\,\mbox{and}
\\[1.5em]
p=3&\Rightarrow 
\xi(\phi)
=\frac{3}{2^{1+\gamma}\alpha}\left[\frac{(\phi+C_3)^2}{1-2\beta(\phi+C_3)^2}\right]^{1-\gamma}\left\{\frac{1}{8(1-\gamma)}+\frac{1}{\left[1-2\beta(\phi+C_3)^2\right]^{p}}\right\}+C.
\end{align}

For the special case in which the attractor for the scalar spectral index is given by $n_s=1-2/N$ i.e., $\gamma=2$, the above reconstructed coupling parameters related to the Gauss-Bonnet term $\xi(\phi)$ for $\beta>0$ are simplified to

\begin{align}
p=1 &\Rightarrow \xi(\phi)
= \frac{3\beta}{4\alpha\,\sinh^2\!\left(\sqrt{2\beta}\phi+C_1\right)}
\left[\sinh^2\!\left(\sqrt{2\beta}\phi+C_1\right)+\frac{7}{8}\right]
+C, \\
p=2 &\Rightarrow \xi(\phi)
= \frac{3\beta}{4\alpha\,\tan^2\!\left(\sqrt{2\beta}\phi+C_2\right)}
\left[\tan^2\!\left(\sqrt{2\beta}\phi+C_2\right)+\frac{7}{8}\right]
+C, \,\,\,\,\,\mbox{and}\\
p=3 &\Rightarrow \xi(\phi)
= \frac{3\left[1-2\beta\left(\phi+C_3\right)^2\right]}
{8\alpha\left(\phi+C_3\right)^2}
\left[\frac{2\beta\left(\phi+C_3\right)^2}{1-2\beta(\phi+C_3)^2}+\frac{7}{8}\right]
+C .
\end{align}


On the other hand, in order to constrain the free parameter of our inflationary  model, we will consider the amplitude of the power spectrum of the scalar perturbations  $A_s$ defined by Eq.(\ref{As1}). In this way, we find that the amplitude of the power spectrum of the scalar perturbations for our model reconstructed becomes the following 
\begin{equation}
A_{s_*}\simeq\left(\frac{H_*}{2\pi}\right)^2\frac{1}{F_*}=\frac{\alpha N_*^{\gamma-1}}{12\pi^2F_*(1+\beta N_*)^p}=\frac{\alpha N_*^{\gamma-1}}{12\pi^2(1+\beta N_*)^p}\,\left(\frac{1}{2\epsilon_{1_*}-\delta_{1_*}-\delta_1\epsilon_{1_*}+\delta_{1_*}\delta_{2_*}+(3/2)\delta_{1_*}^2}\right),\label{Ass}
\end{equation}
where we have considered that the parameter $F$ defined by Eq.(\ref{FF}) can be approximated by $F\simeq 2\epsilon_1-\delta_1-\delta_1\epsilon_1+\delta_1\delta_2+(3/2)\delta_1^2(1+\cdots)= 2\epsilon_1-\delta_1-\delta_1\epsilon_1+\delta_1\delta_2+(3/2)\delta_1^2$ and under the slow-roll approximation $c_s\simeq 1$ and $\nu\simeq 3/2$, see Eq.(\ref{nu}). We also note that in the GR limit, where the coupling parameter 
$\xi\rightarrow 0$, the power spectrum $A_s$ reduces to the standard expression given by $A_s=(H/2\pi)^2(1/2\epsilon_1)$ \cite{Guo:2010jr,Yi:2018gse}.
In addition, in the following the notation ``$*$'' denotes  the epoch at which the cosmological
scale exits the horizon.  

In order to determine a constraint on a parameter from the power spectrum given by Eq.(\ref{Ass}), we find that the integration constant $\alpha$ is fixed to
\begin{equation}
\alpha=A_{s_*}\,\left[\frac{12\pi^2F_*(1+\beta N_*)^p}{ N_*^{\gamma-1}}\right]\label{As2},
\end{equation}
where the function $F_*$ associated with the slow roll parameters is given by 
\begin{align}
F_*
=
\frac{1}{128\,u^{3}N_*^{3}}
\Bigg[
& 16 u^{3-p} N_*^{2}
+2\Big(
u^{1-p}\big[1+(u-1)(1+p)\big]
+16\big[1-\gamma+(u-1)\big(2+(u-1)(1+p-\gamma)-2\gamma\big)\big]
\Big)v
\nonumber\\
&+3uv^{2}N_*
-16uvN_*\big[p\,(u-1)+u\, (1-\gamma)\big]
\Bigg] , 
\end{align}
where variables $u$ and $v$ are defined as
\begin{equation}
u \equiv 1 + \beta N_* , \,\,\,\,\,\,\,\,\mbox{and}\,\,\,\,\,\,\,\,\,\,\,
v \equiv 16(1-\gamma) + u^{\,1-p} + 16 (u-1)(1+p-\gamma)  ,
\end{equation}
respectively.

On the other hand, in relation to the parameter $\beta$ and its observational constraint, it satisfies the  inequality 
$\big[N_*(1+\beta N_*)^p\big]^{-1}<r_*$. Thus, in particular, by  
 considering  the number of $e-$folds  $N_*=60$, we can find different lower bounds on the parameter $\beta$ from the  same values of the parameter $p$ and the tensor to scalar ratio at the crossing time $r_*$. In general, the lower bound of the parameter $\beta$ can be determined from the relation $\beta>[(N_*r_*)^{-1/p}-1]/N_*$.
 In this sense, different  lower bounds on $\beta$
 are summarized in Table 
\ref{tablabetas}, for representative values of $p=1,2,3$ and
$r_*$. Here, we note that, considering the upper bound of the tensor-to-scalar ratio together with  different values of the parameter $p$, the values of the parameter $\beta$ (lower bound) are negative.

\begin{table}[ht]
\centering
\renewcommand{\arraystretch}{1.3}
\begin{tabular}{|c|c|c|c|}
\hline
$r_*$ & $p=1$ & $p=2$ & $p=3$ \\
\hline
$0.039$ & $\beta>-9.54\times 10^{-3}$ & $\beta>-5.77\times 10^{-3}$ & $\beta>-4.12\times 10^{-3}$ \\ \hline
$0.020$ & $\beta>-2.78\times 10^{-3}$ & $\beta>-1.45\times 10^{-3}$ & $\beta>-9.8\times 10^{-4}$ \\ \hline
$0.010$ & $\beta>1.11\times 10^{-2}$  & $\beta>4.85\times 10^{-3}$  & $\beta>3.09\times 10^{-3}$ \\
\hline
\end{tabular}
\caption{Lower bounds on the parameter $\beta$ obtained from 
$\big[N_*(1+\beta N_*)^p\big]^{-1}<r_*$ with $N_*=60$ for different values of $r_*$ and $p$.}
\label{tablabetas}
\end{table}

\begin{figure}[ht!]
    \centering
\includegraphics[width=0.9\linewidth]{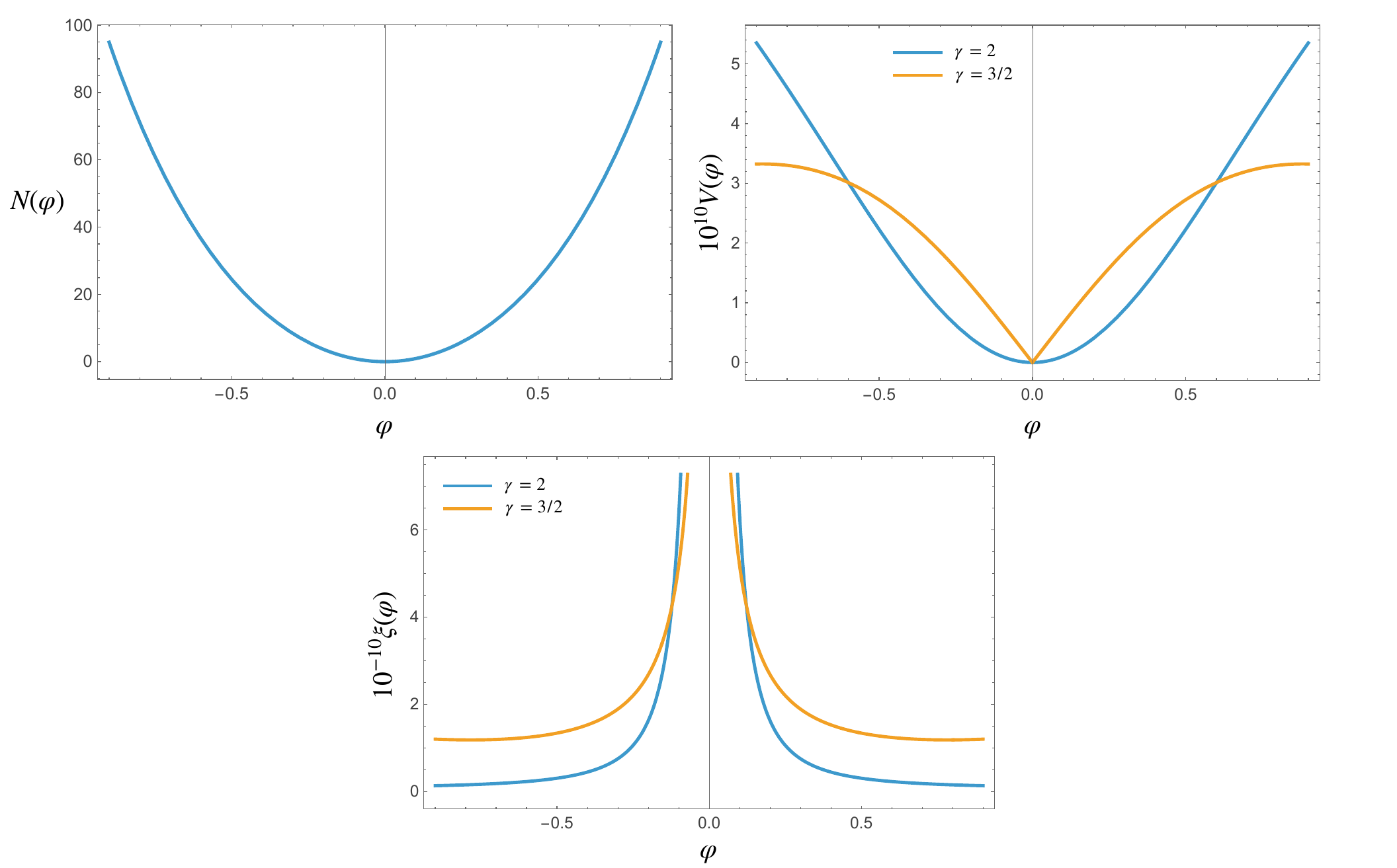}
    \caption{The upper-left panel shows the number of $e$-folds $N$ as a function of the scalar field  $\varphi=\sqrt{2\beta}\phi+C_1$. The upper-right panel displays the reconstructed scalar potential $V(\varphi)$ in terms of $\varphi$. The lower panel shows the reconstructed coupling function associated with the Gauss-Bonnet term as a function of the field $\varphi$. In all panels we have considered the special case in which the parameter $p=1$. In addition, we have used the constant $C=0$ together with  two different values of the parameter $\gamma$ related to the scalar spectral index; $\gamma=3/2$ and $\gamma=2$, respectively.   
    } 
\label{Fig1}
\end{figure}

In Fig.\ref{Fig1}  the upper-left panel shows the number of $e$-folds $N$  as a function of the shifted field $\varphi =\sqrt{2\beta}\phi+C_1$. The upper-right panel shows the reconstructed effective potential as a function of $\varphi$.
The lower  panel shows the reconstructed coupling function associated with the Gauss-Bonnet gravity $\xi(\phi)$ in terms of $\varphi$.
In all panels we consider the special case in which the parameter $p=1$. Also, in the right and lower panels, we assume two different values of the parameter $\gamma$ associated with the attractor $n_s=1-\gamma/N$; $\gamma=3/2$, which is consistent with the ACT data, and $\gamma=2$, which provides a good fit to the Planck data. We also emphasize that the number of $e-$folds $N$ as a function of the scalar field does not depend on the parameter $\gamma$, for this reason we plot a single curve in Fig.\ref{Fig1} (upper-left panel), see Eq.(\ref{Nf3}).
 In addition, we have used the value of the parameter $\alpha$   given by Eq.(\ref{As2}) and for the parameter $\beta$, the positive value $\beta=1.11\times 10^{-2}$, see Table \ref{tablabetas}, when $p=1$. As shown in Fig.\ref{Fig1}, the inflationary dynamics can occur in two distinct branches, corresponding to positive or negative values of the shifted  field $\varphi$. From the  upper-left panel, we see that in both branches, inflation terminates as the field approaches zero, i.e., $N(\varphi\sim0)\simeq 0$.

\section{Conclusions }\label{Conc}

In this work, we have investigated the reconstruction from the cosmological parameters of the background variables during the inflationary epoch,  in the context of the slow-roll approximation, 
within the framework of  Einstein-Gauss-Bonnet gravity. Under a general formalism (in integral form), we have found explicit expressions for the effective potential  and the coupling parameter associated with the Gauss-Bonnet term as  functions of the scalar spectral index $n_s(N)$ and the tensor-to-scalar ratio $r(N)$ in terms of the number of $e-$folds $N$.

As a specific example of the parametrization of the scalar spectral index $n_s(N)$ and the tensor-to-scalar ratio $r(N)$, we have studied the general  attractors forms $n_s=1-\gamma/N$ and $r=1/(N(1+\beta N)^p)$. In particular, for the specific case in which the parameter $\gamma\simeq3/2$ in $N=60$, the scalar spectral index is in agreement  with recent  ACT data and for $\gamma=2$ in $N=60$, the index $n_s$ is consistent with Planck results. Here, we have utilized our general expressions to obtain the reconstruction of the background variables.
In this sense, we have obtained that the expressions for the effective potential and the coupling parameter of Gauss-Bonnet gravity in terms of the number of $e-$folds are given by Eqs.(\ref{VNN}) and (\ref{xi1}), respectively.

To reconstruct the background variables in terms of the scalar field  within the framework of Einstein-Gauss-Bonnet gravity, we have found an analytical expression for the number of $e-$folds $N$ and the scalar field $\phi$, see  Eq.(\ref{Nf3}).   
Here, we have obtained  that the relation between the number of $e-$folds $N$ and the scalar field defined by Eq.(\ref{Nf3}) does not depend on the parameter $\gamma$ associated to the scalar spectral index $n_s$. 

Hence, the reconstruction of the effective potential and the coupling function associated with the Gauss-Bonnet term are given by Eqs.(\ref{Pot1}) and (\ref{Xia}), respectively. In addition, we have considered some values of the parameter $p$ ($p=1,2$ and 3) related to the tensor-to-scalar ratio $r(N)$ given  by Eq.(\ref{rAtr}), to  visualize the form of the different reconstructed effective potentials $V(\phi)$ and coupling functions $\xi(\phi)$ for these values of $p$.
Also, we have found that  the reconstruction based on  observational parameters  $n_s(N)$ and $r(N)$ shows that $V(\phi)\neq 1/\xi(\phi)$, in contrast to the assumption 
 adopted in the literature when studying  the inflationary and present universe within the framework of Einstein-Gauss-Bonnet gravity.

Additionally, we have determined a constraint on the free parameter $\alpha$ associated with an integration constant (Eq.(\ref{As2})), using the power spectrum defined by Eq.(\ref{Ass}) within the framework of Einstein-Gauss-Bonnet gravity. Moreover, for the special case $p=1$ and considering  the different values of the space-parameter ($\alpha,\beta$), we have plotted in 
Fig.\ref{Fig1}, the reconstruction of the effective potential $V(\varphi)$ and the coupling function $\xi(\varphi)$  in terms of the shifted field $\varphi=\sqrt{2\beta}\,\phi+C_1$.  In this figure, we have considered two values for the parameter $\gamma$ associated to the scalar spectral index; $\gamma=2$, which is supported by Planck data and $\gamma=3/2$ in  light of ACT results.

Finally, in this article, we have not addressed the reconstruction of the background variables, in the context of Einstein-Gauss-Bonnet gravity  for other parametrizations of  the scalar spectral index  and the tensor-to-scalar ratio  in terms of the number of $e-$folds. We hope to revisit  this point in the near future.

\bibliography{bio}

\end{document}